\documentclass[aps,prb,superscriptaddress,showpacs,amsmath,amssymb,letterpaper,twocolumn]{revtex4}
\usepackage{graphicx}
\usepackage{wasysym}
\usepackage{epsfig,color,verbatim} 
\usepackage{hyperref}

\begin{document}
\newcommand{\boldnabla}{\mbox{\boldmath$\nabla$}}
\newcommand{\boldrho}{\mbox{\boldmath$\rho$}}

\title{On applicability of inhomogeneous diffusion approach to localized transport through disordered waveguides}
\author{Pauf Neupane and Alexey G. Yamilov\footnote{e-mail:yamilov@mst.edu}}
\affiliation{Department of Physics, Missouri University of Science \& Technology, Rolla, Missouri 65409, USA}

\date{\today}
\begin{abstract}
In this work we show analytically and numerically that wave transport through random waveguides can be modeled as a diffusion with an inhomogeneous diffusion coefficient (IDC). In localized regime, IDC retains the memory of the source position. In an absorbing random medium, IDC becomes independent of the source.
\pacs{42.25.Dd,42.25.Bs,72.15.Rn}
\end{abstract}
\maketitle 

\section{Introduction\label{sec:intro}}

Diffusive description of wave transport random  medium has a long history~\cite{1953_Morse}. This macroscopic approach describes the ensemble-averaged intensity of the wave on scales longer than transport mean free path $\ell$. As such, it has a tremendous practical advantage compared to the direct solution of wave equation for each statistical realization of disorder and subsequent averaging over the ensemble of solutions. 

Diffusion coefficient can become renormalized~\cite{1980_Vollhardt_Wolfle} due to wave localization phenomenon~\cite{2009_Lagendijk_PT}. In three-dimensions for sufficiently strong disorder, the diffusion vanishes for an infinitely large system~\cite{1958_Anderson}. In practice, however, one deals with transport through finite systems. In process of adapting self-consistent theory (SCT) of localization~\cite{1980_Vollhardt_Wolfle,1993_Kroha_self_consistent} and super-symmetric (SUSY) theories~\cite{1997_Efetov_SUSY_book}, it became clear~\cite{2000_van_Tiggelen,2008_Tian} that the diffusive-like description can also be applied to the {\it finite} systems that exhibit localized transport, in particular, to the low dimensional systems. In such description the diffusion coefficient becomes inhomogeneous in space and dependent on system size~\cite{2008_Cherroret,2010_Payne_PRL,2010_Tian_PRL} and geometry~\cite{2014_Sarma_Taper_Waveguides}. In quasi-1D or 1D lossless medium both SCT and SUSY lead to the following equation for the ensemble-averaged intensity $\langle I\left(z,z^\prime\right)\rangle$ in the presence of a point source $J_0$ at $z^\prime$
\begin{equation}
-(d/dz) D(z)d\langle I\left(z,z^\prime\right)\rangle/dz=J_0\delta\left(z-z^\prime\right)
\label{eq:diffusion_equation}
\end{equation}
Diffusion of this kind results leads to a highly unusual macroscopic transport~\cite{2013_Tian_review}. We stress that the medium itself, i.e. the density of scatterers, is uniform and that the inhomogeneous diffusion is brought about by non-local wave interference effects.

To date, the studies of inhomogeneous wave diffusion has concentrated on geometry where a wave is incident onto the random medium from an outside (free space) region~\cite{2000_van_Tiggelen,2008_Tian,2008_Cherroret,2008_van_Tiggelen_Nature,2009_Genack_PRB,2010_Payne_PRL,2010_Tian_PRL,2014_Yamilov_Dofz_experiment,2014_Sarma_Taper_Waveguides,2014_Scheffold_Waveguide}. Inhomegeneous wave diffusion description was successful in describing the light intensity under various measurement conditions, see Ref.~\cite{2013_Tian_review} for a review. Ensemble-averaged intensity, however, is only the first step in characterizing wave transport in random media. Indeed, the second order statistical quantities, such as fluctuations or correlations, become important at the onset of Anderson localization~\cite{2000_chabanov_nature,2005_Genack_review}, they require the knowledge of the green function of the diffusion equation, e.g. $\langle I\left(z,z^\prime\right)\rangle$ with arbitrary $z^\prime$~\cite{,1994_Berkovits_Feng,1999_van_Rossum,2007_Akkermans_book}. 

In this work we test the applicability of Eq.~\ref{eq:diffusion_equation} with an arbitrary position of the source. We show analytically that for $z^\prime\neq 0$, the diffusion equation is applicable, however inhomogeneous diffusion coefficient (IDC) $D(z)$ acquires the dependence on the position of the source $z^\prime$. We derive an closed form analytic expression for $D\left( z,z^\prime\right)$ and verify it with ab-initio numerical simulations. We show that $D\left( z,z^\prime\right)$ is reduced to the known result~\cite{2010_Tian_PRL} for $z^\prime\rightarrow 0$, i.e. for wave incident from the outside region. We demonstrate that when absorption, unavoidable in an experiment, is present in the system, $D\left( z,z^\prime\right)$ looses its dependence on source position and that it can be determined using self-consistent theory~\cite{2008_Cherroret,2010_Payne_PRL}. 

This paper is organized as follows. In Sec.~\ref{sec:outside} we obtain an analytical expression for IDC that describes wave transport in a single mode waveguide with an external source. We verify the applicability of the result with numerical simulations. In Sec.~\ref{sec:inside}, we demonstrate analytically and confirm numerically that IDC depends on the position of the source inside random medium. The effect of absorption is accounted for in Sec.~\ref{sec:absorption}, where we show that IDC looses is position dependence for a sufficiently strong absorption in the system.

\section {Inhomogeneous diffusion in passive system with an external source\label{sec:outside}}

We consider one-dimensional random medium occupying $0\leq z\leq L$ region. Propagation of a scalar wave $E(z)$ is described by 
\begin{equation}
\frac{d^2E(z)}{dz^2}+k^2\left[1+\epsilon(z)\right]E(z)=0,
\label{eq:wave_eq}
\end{equation}
here $k=2\pi/\lambda$ is wave number and $\epsilon(z)$ is a random process. For a wave with unit amplitude incident from the left, the boundary conditions can be expressed in terms of reflection $r$ and transmission $t$ coefficients as
\begin{eqnarray}
E(z)&=&e^{ikz}+r\ e^{-ikz},\,\,\,z<0\nonumber \\
E(z)&=&te^{ikz},\,\,\,z>L
\label{eq:wave_eq_bc}
\end{eqnarray}
We are interested in obtaining a closed form expression for the intensity $\langle I(z)\rangle\equiv\langle\left|E(z)\right|^2\rangle$ averaged over the ensemble of random processes $\epsilon(z)$. Indeed, inhomogeneous diffusion coefficient $D(z)$ can be found from the Fick's law
\begin{equation}
\langle J(z)\rangle=-D(z)d\langle I(z)\rangle/dz,
\label{eq:Ficks_law}
\end{equation}
where $\langle J(z)\rangle$ is the flux. In passive random medium the flux is conserved during propagation and, thus, can be found from the boundary conditions. Indeed, the fraction of the flux propagating in positive ($+$) or negative ($-$) direction can be expressed as~\cite{1953_Morse} 
\begin{equation}
\langle J^{(\pm)}(z)\rangle=v\langle I(z)\rangle\mp \left(D(z)/2\right)\ d\langle I(z)\rangle/dz
\label{eq:Jpm}
\end{equation}
where $v$ is the wave speed. The right boundary $\langle J^{(-)}(z)\rangle$ vanishes so $\langle J(z)\rangle=\langle J^{(+)}(L)\rangle=2v\langle I(L)\rangle$. Substituting this expression into Eq.~(\ref{eq:Ficks_law}) we obtain
\begin{equation}
D(z)=-2v\langle I(L)\rangle/\left(d\langle I(z)\rangle/dz\right).
\label{eq:Dofz}
\end{equation}
Therefore, finding IDC requires the knowledge of (only) $\langle I(z)\rangle$ for the problem defined by Eqs.~(\ref{eq:wave_eq},\ref{eq:wave_eq_bc}). Such solution has been obtained in Refs.~\cite{1969_Gazaryan,1973_Papanicolaou_Resonant_Transmission,1992_Klyatskin_Resonant_Tunneling}. The common theme is these studies is to relate the statistical property of the wave field inside the medium to those at the boundary~\ref{eq:wave_eq_bc} where it can be obtained using the limiting theorems. Such approach is in spirit of the well known self-embedding method~\cite{1987_Rammal_Doucot_Imbedding}. We will assume that $\epsilon(z)$ is a delta-correlated Gaussian process with $\langle\epsilon(z)\rangle$ and $\langle\epsilon(z)\epsilon(z^\prime)\rangle=a\delta(z-z^\prime)$. Under these conditions the solution for the ensemble-averaged intensity is obtained in the form~\cite{1969_Gazaryan,1973_Papanicolaou_Resonant_Transmission,1992_Klyatskin_Resonant_Tunneling}
\begin{widetext}
\begin{equation}
\langle I(z)\rangle =1-\sqrt{\frac{\xi} {\pi L}}\int_{-\infty}^{\infty}
\exp\left[ -\frac{(\zeta-(z-L/2)/\xi)^2}{L/\xi} \right]\\
\left(\tanh(\zeta)+\frac{\zeta}{\cosh(\zeta)^2}\right)d\zeta,
\label{eq:Iofz}
\end{equation}
\end{widetext}
where we introduced localization length as $\xi^{-1}=ak^2/2$. Substitution of Eq.~(\ref{eq:Iofz}) into Eq.~(\ref{eq:Dofz}) gives us the analytical expression for IDC. 

\begin{figure}
\centering
\includegraphics[width=3in]{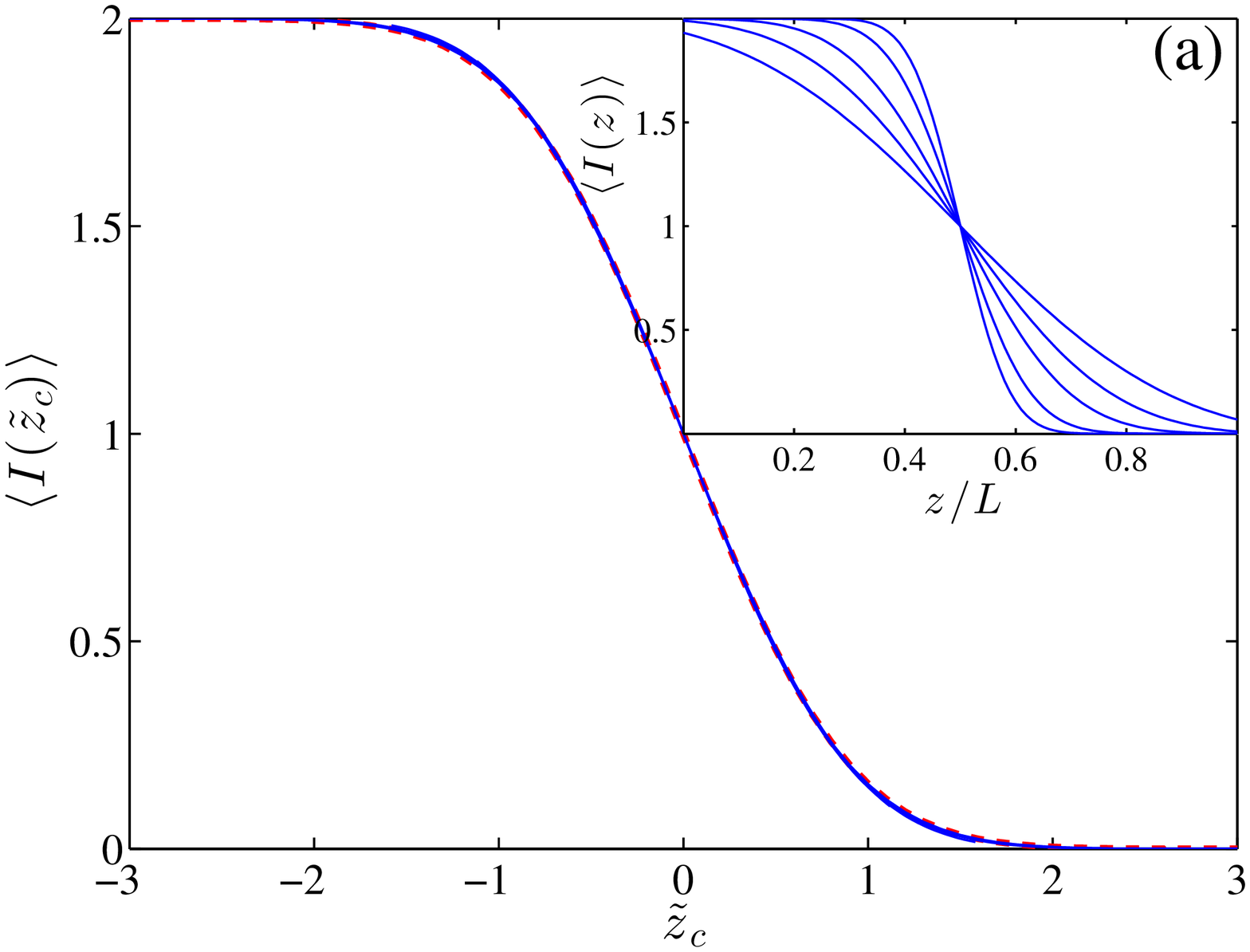}
\includegraphics[width=3in]{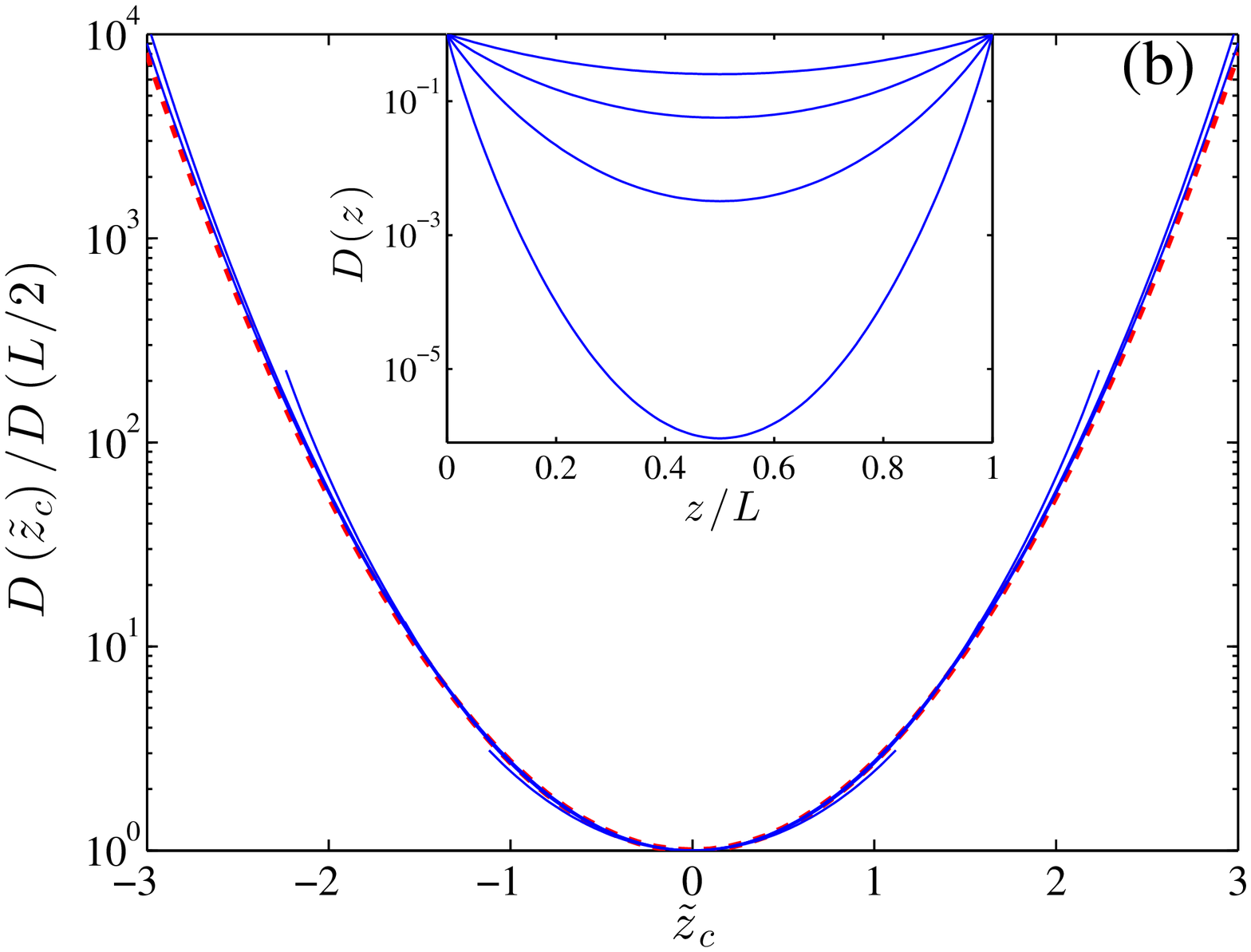}
\caption{(a) Spatial distribution of ensemble-averaged intensity $\langle I(z)\rangle$ from Eq.~(\ref{eq:Iofz}) (blue lines) without scaling (inset) for systems with $L/\xi=5,10,20,50,100$ (left to right). The main panel presents the same data in terms of scaled coordinate $\tilde{z}_c=(z-L/2)/\sqrt{L\xi}$; the approximate expression given by Eq.~(\ref{eq:Iofz_simplified}) is shown with thick dashed line. (b) Inhomogeneous diffusion coefficient found from Eq.~(\ref{eq:Dofz}) for systems with $L/\xi=5,10,20,50$ without (inset) and with (main panel) scaling. Thick dashed line is found from asymptotic $L\gg\xi$ expression in Eq.~(\ref{eq:Dofz_simplified}).}
\label{fig:passive_1D}
\end{figure}

A compact expressions for both $\langle I(z)\rangle$ and $D(z)$ can be obtained when $L\gg\xi$. In this limit, the expression in parentheses in the integrand of Eq.~(\ref{eq:Iofz}) can be approximated with a step function $h(\zeta)$ and the integral can be computed in terms of the error function ${\rm erf}(x)$:
\begin{equation}
\langle I(z)\rangle \simeq 1-{\rm erf}\left[ \tilde{z}_c \right]
\label{eq:Iofz_simplified}
\end{equation}
with scaling parameter $\tilde{z}_c=(z-L/2)/\sqrt{L\xi}$ as the argument. In Fig.~\ref{fig:passive_1D}a, $\langle I(z)\rangle$ is plotted for $L/\xi$ = 5, 10, 20, 50 and 100 with and without scaling $z$-coordinate. We confirm that Eq.~(\ref{eq:Iofz_simplified}) approximates the exact expression Eq.~(\ref{eq:Iofz}) well. We note, that such distribution has been observed in numerical simulations of energy deposition in wavefront shaping in random medium~\cite{2014_Cheng_Wavefront_Shaping_Energy_Deposition}.

Asymptotic expression IDC in $L\gg\xi$ is obtained by substituting Eq.~(\ref{eq:Iofz_simplified}) into Eq.~(\ref{eq:Dofz}):
\begin{equation}
D(z)\simeq D_0\exp\left[ \tilde{z}_c^2 -L/4\xi\right].
\label{eq:Dofz_simplified}
\end{equation}
Here $D_0=v\ell$ is unrenormalized value of diffusion coefficient in terms of transport mean free path $\ell=2\xi$. Fig.~\ref{fig:passive_1D}b confirms the universality of IDC inside 1D passive random medium in terms of the scaling parameter $\tilde{z}_c$.  Eq.~(\ref{eq:Dofz_simplified}) agrees with that derived in super-symmetric theory of Ref.~\cite{2010_Tian_PRL} for quasi-1D geometry (a multimode waveguide). 

\section {Inhomogeneous diffusion in passive system with an internal source\label{sec:inside}}

For the source located inside random medium, Eq.~(\ref{eq:wave_eq}) is modified to include a point source at $z^\prime$:
\begin{equation}
\frac{d^2E(z,z^\prime)}{dz^2}+k^2\left[1+\epsilon(z)\right]E(z,z^\prime)=\delta(z-z^\prime),
\label{eq:wave_eq_inside}
\end{equation}
whereas the boundary conditions in Eq.~(\ref{eq:wave_eq_bc}) are replaced with the outgoing wave conditions at both ends of the waveguide 
\begin{eqnarray}
E(z)&=&t_1\ e^{-ikz},\,\,\,z<0\nonumber \\
E(z)&=&t_2\ e^{ikz},\,\,\,z>L
\label{eq:wave_eq_bc_inside}
\end{eqnarray}
Under these conditions, the flux inside the medium is a piece-wise constant function with a jump at the position of the source $z^\prime$. The values of $\langle J(z,z^\prime)\rangle$ for $z<z^\prime$ and $z>z^\prime$ can be determined by applying Eq.~(\ref{eq:Jpm}) at $z=0$ and $z=L$ respectively. We find $\langle J(z<z^\prime)\rangle=-2v\langle I(0,z^\prime)\rangle$ and $\langle J(z>z^\prime)\rangle=2v\langle I(L,z^\prime)\rangle$, where $\langle I(z,z^\prime)\rangle=\langle\left| E(z,z^\prime)\right|^2\rangle$ and $E(z,z^\prime)$ is the solution of Eq.~(\ref{eq:wave_eq_inside}) with boundary conditions in Eq.~(\ref{eq:wave_eq_bc_inside}). Therefore, IDC can be written based on Fick's law Eq.~(\ref{eq:Ficks_law}) as
\begin{equation}
D(z,z^\prime)=-2v\left(d\langle I(z)\rangle/dz\right)^{-1}
\left\{ \begin{array}{ll}
        -\langle I(0,z^\prime)\rangle & \mbox{, $z<z^\prime$};\\
         \langle I(L,z^\prime)\rangle & \mbox{, $z>z^\prime$}.\end{array} \right.
\label{eq:Dofz_inside}
\end{equation} 
As in Sec.~\ref{sec:outside}, the above expression for IDC requires the knowledge of the ensemble-averaged intensity $\langle I(z,z^\prime)\rangle$. The latter has been obtained in Refs.~\cite{1974_Papanicolaou_Resonant_Transmission,1992_Klyatskin_Resonant_Tunneling} in the form
\begin{widetext}
\begin{equation}
\begin{split}
\langle I(z,z^\prime)\rangle=\pi\exp\left[\frac{3 L}{4\xi}-\frac{|z-z^\prime|}{\xi}\right]\int_{-\infty}^\infty d\mu \exp\left[-\frac{\mu^2L}{\xi} \right]
\frac{\sinh(\pi\mu)}{\mu\cosh(\pi\mu)^2}[(\mu^2+\frac{1}{4})\cos(2\mu\frac{(z+z^\prime-L)}{\xi})\\
+(\mu^2-\frac{1}{4})\cos(2\mu\frac{(L- |z-z^\prime|)}{\xi}) +\mu\sin(2\mu\frac{(L-|z-z^\prime|)}{\xi} )].
\end{split}
\label{eq:Iofzzp}
\end{equation}
\end{widetext}
Substituting this expression into Eq.~(\ref{eq:Dofz_inside}) gives us the final result. 

\begin{figure}
\centering
\includegraphics[width=3.5in]{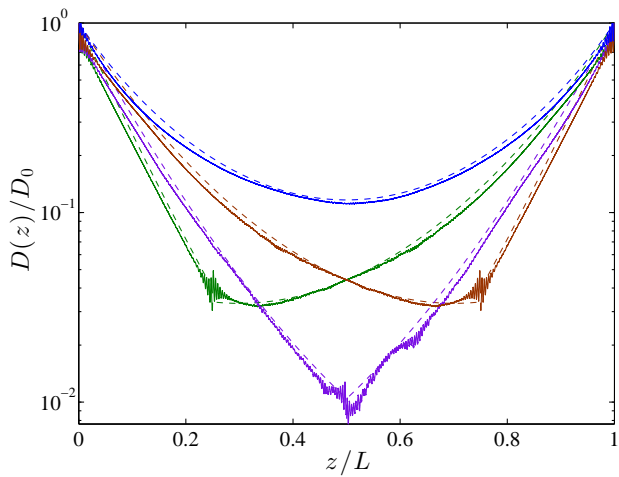}
\caption{\label{fig:inside} (Color online) Inhomogeneous diffusion coefficient inside 1D passive random media of $L=7.6\xi$ computed for four source positions $z^\prime=0$ (upper curves), $L/4$, $L/2$ and $3L/4$. The latter three curves have cusps at the position of the source.  Dashed lines were found by substituting the analytical result Eq.~(\ref{eq:Iofzzp}) into Eq.~(\ref{eq:Dofz_inside}). The solid lines were obtained numerically.}
\end{figure}

We make the following observations. First of all, in the limit of $z^\prime=0$, we recover the result for an external source found in the previous section. Indeed, Eq.~(\ref{eq:Iofzzp}) can be shown~\cite{1969_Gazaryan} to reduce to Eq.~(\ref{eq:Iofz}). Second, unlike Eq.~(\ref{eq:diffusion_equation}), IDC $D(z,z^\prime)$ depends on the source position $z^\prime$. In Fig.~\ref{fig:inside} we evaluated Eq.~(\ref{eq:Dofz_inside}) for $L/\xi=7.6$ and four values of $z^\prime$: $0$ (outside source), $L/4$, $L/2$ and $3L/4$. Indeed, IDC shows strong dependence on the position of the source. We note that $D(z,0)$ is always greater than $D(z,z^\prime>0)$ with the minimum value at the middle of the sample for $z^\prime=L/2$. 

We verified the above results with numerical simulations of wave with $k=1.45$ propagating normally through a stack of alternating dielectric slabs with dielectric constants $\epsilon_1=1$ and $\epsilon_2=1.2$. The widths of the stacks of the first kind is distributed uniformly in the interval $d_1\in (0.9,1.1)$, while the width of the others is being kept constant $d_2=1$. The propagation of the waves in the system consists of free propagation in the slabs and scattering at the interfaces, where the boundary conditions should be satisfied. It can be described using the transfer-matrix formalism, c.f. Ref.~\cite{1999_yamilov_selfembed,2001_Deych_sps_abs}. We computed $J(z)=-kc{\rm Im}\left[E(z)dE(z)/dz\right]$ and $I(z)=(k^2/2)\left|E(z)\right|^2+(1/2)\left|dE(z)/dz\right|^2$ in a system with $N=8\times 10^3$ layers numerically and averaged over $10^8$ disorder realizations. IDC was found from Fick's law Eq.~(\ref{eq:Ficks_law}). The results reported in Fig.~\ref{fig:inside} (sold lines) agree with the analytical expression (dashed lines). 

\section {Effect of absorption\label{sec:absorption}}

Absorption is inevitable in optical experiments. The effect of absorption is to suppress  resonant tunneling of the wave~\cite{2010_Tian_PRL}, thereby to increase IDC~\cite{2013_Yamilov_Localization_with_Absorption,2013_Tian_diffusion_with_abs}. In this section we perform numerical analysis of the effect of absorption on IDC for an internal source.

Modeling absorbing random medium is accomplished by adding a constant imaginary part to $\epsilon(z)$ in Eq.~(\ref{eq:wave_eq_inside}) as $\epsilon(z)\rightarrow\epsilon(z)+i\gamma$. Addition of the loss results in an extra term $\xi_a^{-2}\langle I(z,z^\prime)\rangle$ in the inhomogeneous diffusion equation. Absorption length $\xi_a$ can be obtained for a given value of $\gamma$ from the continuity condition $d\langle J(z,z^\prime)\rangle/dz=\langle I(z,z^\prime)\rangle/\tau_a$ where $\tau_a=\xi_a^2/D_0$~\cite{2013_Yamilov_Localization_with_Absorption}.

\begin{figure}
\centering
\includegraphics[width=3.5in]{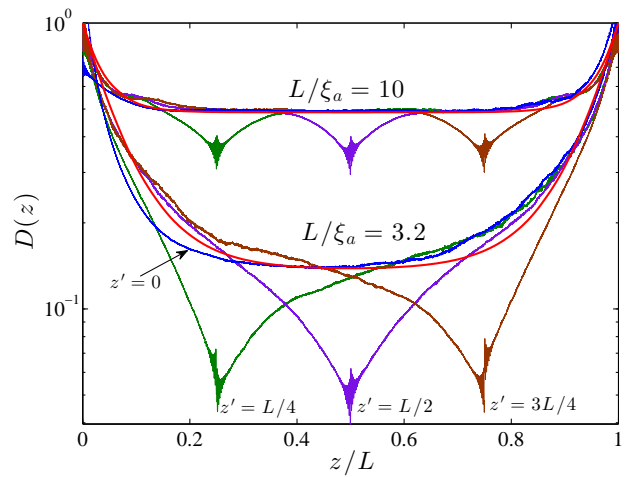}
\caption{\label{fig:absorption} (Color online) Inhomogeneous diffusion coefficient in absorbing random medium with $L/\xi=15.2$. The lower set of curves corresponds to absorption $L/\xi_a=3.2$ and four positions of the source. Thick red line is the prediction of SCT from Eq.~(\ref{eq:diffusion_equation_SCT},\ref{eq:Dofz_SCT}). For stronger absorption $L/\xi_a=10$ (upper set), SCT predicts IDC well.}
\end{figure}

Figure~\ref{fig:absorption} shows IDC obtained numerically for the model in Sec.~\ref{sec:inside}. We choose number of layers in a stack $N=1.6\times 10^4$ and two values of $\gamma$: $10^{-5}$ and $10^{-4}$. These parameters give $L/\xi_{a0}$: $3.2$ and $10$ respectively. Similar to the passive (non-absorbing) case in Sec.~\ref{sec:inside}, IDC clearly shows dependence on the position of the source inside the medium ($z^\prime=0$, $L/4$, $L/2$ and $3L/4$ are shown), it has a cusp feature at $z^\prime$. However, closer inspection of Fig.~\ref{fig:absorption} shows that dependence on source position is strongly suppressed at large $L/\xi_a$; in this limit $\xi_a$ becomes comparable to the localization length $\xi$. 

Performing computationally expensive numerical simulations is not practical, in particular, in higher dimensional systems. Self-consistent theory~\cite{2000_van_Tiggelen,2008_Cherroret,2010_Payne_PRL} has been successful in providing a good prediction for IDC for systems with $L/\xi$ not too large~\cite{2010_Tian_PRL}, and it was shown~\cite{2013_Yamilov_Localization_with_Absorption} to be accurate in the absorbing systems. In all previous works, external source has been considered. Here, we computed the prediction of the self-consistent theory for systems with different amount of absorption. Self-consistency condition relates IDC to the return probability $\langle I(z,z)\rangle$ as 
\begin{eqnarray}
&-\frac{d}{dz}D^{(SCT)}(z) \frac{d\langle I\left(z,z^\prime\right)\rangle}{dz}+\frac{\langle I\left(z,z^\prime\right)\rangle}{\xi_a^2}=J_0\delta\left(z-z^\prime\right)\label{eq:diffusion_equation_SCT} \\
&D_0/D^{(SCT)}(z)=1+v\langle I(z,z)\rangle.
\label{eq:Dofz_SCT}
\end{eqnarray}
These equations form a closed set, sufficient for finding $D^{(SCT)}(z)$. The solution of Eqs.~(\ref{eq:diffusion_equation_SCT},\ref{eq:Dofz_SCT}) is shown with a thick solid line in Fig.~\ref{fig:absorption}. We find that in presence of sufficiently strong absorption SCT makes adequate prediction for IDC even for internal sources.

\section {Conclusion\label{sec:conclusion}}

In this work we investigated the applicability of the inhomogeneous diffusion approach to describing wave transport in random medium. We have shown analytically and numerically that in the regime of localized transport $L/\xi$, the inhomogeneous diffusion coefficient can be defined through Fick's law. The benefit of such approach, is that it allows one to obtain the ensemble-averaged value of intensity without the need to perform statistical averaging.

Our analysis shows that IDC exhibits a significant dependence on the source position $z^\prime$. Such dependence has not been discussed before. This is because previous studies have concentrated on the common experimental arrangement -- the incident wave is impingent onto the sample from the outside. Present study of the IDC with an internal source is of practical interest for a number of reasons. First of all, solution $\langle I(z,z^\prime)\rangle$ of the diffusion equation with IDC with an internal source $z^\prime$ is the green function which can be used to define second order statistics (e.g. fluctuation, correlations) of wave transport. Secondly, Fick's law with IDC in the form of $D(z,z^\prime)$ points to a highly unconventional type of diffusion in the localized systems. The spatial dependence of IDC has been shown~\cite{2013_Tian_review} to exhibit an unusual macroscopic transport behavior. The additional dependence on the source position found in this work may necessitate completely new {\it non-local approach} to transport. Accurate description of wave transport through random media would inform the studies of limits of wavefront shaping~\cite{2012_Mosk_SLM_review} with applications, in particular, in the field of biological imaging~\cite{2015_Yu_Wavefront_Shaping_Review}.

\section {Acknowledgment\label{sec:acknowledgement}}
Financial support was provided by National Science Foundation under grants No. DMR-1205223.

\bibliographystyle{apsrev4-1}
\bibliography{Dofz_analytical.bbl}

\end{document}